\documentclass[twocolumn,showpacs,preprintnumbers,amsmath,amssymb]{revtex4}

\usepackage{graphicx}
\usepackage{dcolumn}
\usepackage{bm}
\usepackage{epsfig}

\begin{document}

\title{A note on quantum error correction with continuous variables}


\author{Peter van Loock}
\email{pvanloock@optik.uni-erlangen.de}
\affiliation{Optical Quantum Information Theory Group,
Institute of Theoretical Physics I and Max-Planck Research Group,
Institute of Optics, Information and Photonics,
Universit\"{a}t Erlangen-N\"{u}rnberg, Staudtstr. 7/B2,
91058 Erlangen, Germany}

\begin{abstract}
We demonstrate that continuous-variable quantum error correction
based on Gaussian ancilla states and Gaussian operations
(for encoding, syndrome extraction, and recovery) can be very useful
to suppress the effect of non-Gaussian error channels.
For a certain class of stochastic error models,
reminiscent of those typically considered in the qubit case,
quantum error correction codes designed for
single-channel errors may enhance the transfer fidelities
even when errors occur in every channel employed for transmitting the encoded state.
In fact, in this case, the error-correcting capability of
the continuous-variable scheme turns out to be
higher than that of its discrete-variable analogues.
\end{abstract}


\maketitle

\section{Introduction}

Quantum error correction is an essential tool
in order to protect fragile quantum states whenever
they are subject to errors in a quantum computation protocol,
including unwanted decoherence effects due to interactions
with the environment \cite{NielsenChuang}.
In the case of systems described by discrete quantum variables
such as qubits, the theory of quantum error correction codes (QECC)
is very advanced. Various codes have been proposed to correct
arbitrary types of errors. In these codes, a signal state is
encoded into a larger Hilbert space, typically including additional
ancilla systems, resulting in an encoded, multipartite entangled state.
Local errors on some subsystems can then be corrected,
as the quantum information remains undisturbed in the global
structure of the total encoded state.
The simplest qubit QECC
are designed such that only single-qubit errors can be corrected,
i.e., errors occurring in a single channel, assuming that the
subsystems of the encoded state are sent through individual,
independent channels. Provided the error probabilities of
such single-channel errors are below a certain threshold
(and hence, multiple-channel errors are highly unlikely),
a protocol based upon QECC leads to a better performance than
an unencoded scheme.

In the case of systems described by continuous quantum variables,
the situation is not so clear. Although there are proposals of QECC
in the regime of continuous variables
\cite{LloydSlotine,Braunstein1,Braunstein2}, the applicability
of these codes, and to what extent they are useful, has not been fully
understood. For example, encoding an {\it arbitrary} signal state into
a nine-mode wavepacket code would enable one to correct an {\it arbitrary}
single-mode error, including Gaussian or non-Gaussian errors
(where a Gaussian error maps a Gaussian state back to a Gaussian state).
The whole protocol would nonetheless rely upon only Gaussian
ancilla states (squeezed states) and Gaussian operations
(beam splitter transformations, homodyne detection, and feedforward;
online squeezing transformations are not needed
\cite{Braunstein2}).
This would lead to very efficient implementations of QECC.
However, similar to the qubit case, do these codes
also improve the performance of quantum information protocols
in the presence of independent, {\it multiple-channel} errors?
Or do we have to strictly (and rather artificially) assume
that only a single mode in a single channel is affected by an error,
while all the other channels are ideal?
The answer to this question will certainly depend on the
error (or noise) model taken into account. In fact, quite intuitively,
the transmission of an optical mode in a Gaussian state
subject to photon loss cannot be enhanced using the Gaussian QECC,
as the losses would occur in every channel of the encoded state
and the corresponding, realistic noise model would lack the stochastic
nature of the most common qubit channels.
More generally, it has been proven that Gaussian QECC
are not useful to protect Gaussian states against Gaussian errors
\cite{Nisetqph,footnoteGKP}.

Here, we will focus on a simple Gaussian three-mode repetition code
which can correct
arbitrary ``$x$-errors'', i.e., $x$-displacements and any other errors
decomposable into $x$-displacements (including non-Gaussian errors).
A code for correcting arbitrary errors including non-commuting
$x$ and $p$-errors is obtainable, for instance, by concatenating the
three-mode code into a nine-mode code \cite{Braunstein2}.
We find that for a certain class of (non-Gaussian)
error models, the deterministic three-mode QECC protocol enables one to achieve
a significant improvement of fidelity compared to the direct
transmission of the signal state. Such error models may describe,
for instance, free-space channels with
atmospheric fluctuations causing beam jitter,
as considered recently for various non-deterministic distillation
protocols \cite{Heersink,NisetPRL08,Dong} (see also \cite{Schnabel}).

\section{Stochastic error models}

Let us consider the following error model. The input state described
by the Wigner function $W_{\rm in}$ is transformed into a new state
$W_{\rm error}$
with probability $\gamma$; it remains unchanged with probability
$1-\gamma$. Thus, we have
\begin{eqnarray}\label{errormodel}
W_{\rm out}(x,p) = (1-\gamma)  W_{\rm in}(x,p) + \gamma W_{\rm error}(x,p)\,.
\end{eqnarray}
In general, the Wigner functions $W_{\rm in}$ and $W_{\rm error}$
may describe arbitrary quantum states.
Note that even in the case of two Gaussian states,
$W_{\rm in}$ and also $W_{\rm error}$,
the resulting state $W_{\rm out}$ is no longer Gaussian.
Thus, this channel model describes a certain, simple form
of non-Gaussian errors.
It is a generalization of the
``erasure'' channel considered in Ref.~\cite{NisetPRL08}, where
a coherent-state input is displaced to a vacuum state
with an error probability of $\gamma$; otherwise the coherent state
leaves the channel untouched.

Similarly, as an example, we will now consider a coherent-state input,
$|\bar\alpha_1\rangle = |\bar x_1 + i \bar p_1\rangle$,
described by the Wigner function,
\begin{eqnarray}\label{input}
W_{\rm in}(x_1,p_1) = \frac{2}{\pi} \exp[-2 (x_1 - \bar x_1)^2
-2 (p_1 - \bar p_1)^2]\,.
\end{eqnarray}
Moreover, for simplicity, we assume that the effect of the error
is just an $x$-displacement by $\bar x_2$ such that
\begin{eqnarray}\label{output}
W_{\rm error}(x_1,p_1) = W_{\rm in}(x_1-\bar x_2,p_1)\,.
\end{eqnarray}
Note that more general errors, including non-Gaussian $x$-errors,
could be considered as well.
The sign of the displacement error shall be fixed and known,
e.g., without loss of generality, $\bar x_2 > 0$.

\section{Encoding and transmission}

Now in order to encode the input state, we use
two ancilla modes, each in a single-mode $x$-squeezed vacuum state, represented by
\begin{eqnarray}\label{ancilla}
W_{\rm anc}(x_k,p_k) = \frac{2}{\pi} \exp[-2 e^{+2 r} x_k^2
-2 e^{-2 r} p_k^2]\,,
\end{eqnarray}
with squeezing parameter $r$ and $k=2,3$.
The total three-mode state before encoding is
\begin{eqnarray}\label{totalbefore}
W(\alpha_1,\alpha_2,\alpha_3) = W_{\rm in}(x_1,p_1)
W_{\rm anc}(x_2,p_2)W_{\rm anc}(x_3,p_3),
\end{eqnarray}
with $\alpha_j = x_j + i p_j$, $j=1,2,3$. The encoding may be achieved
by applying a ``tritter'', i.e., a sequence of two beam splitters with
transmittances $1:2$ and $1:1$. The total, encoded state will be an
entangled three-mode Gaussian state with Wigner function,
\begin{eqnarray}\label{totalafter}
&&W_{\rm enc}(\alpha_1,\alpha_2,\alpha_3) =
\left(\frac{2}{\pi}\right)^3 \\
&&\times
\exp\Big\{-2 \Big[\frac{1}{\sqrt{3}} \Big(x_1 + x_2 + x_3\Big) - \bar x_1\Big]^2
\nonumber\\
&&\quad\quad\quad\,\,\,
-\frac{2}{3} e^{-2 r} \Big[(p_1-p_2)^2 + (p_2-p_3)^2 + (p_1-p_3)^2\Big]
\nonumber\\
&&\quad\quad\quad\,\,\,
-2 \Big[\frac{1}{\sqrt{3}} \Big(p_1 + p_2 + p_3\Big) - \bar p_1\Big]^2
\nonumber\\
&&\quad\quad\quad\,\,\,
-\frac{2}{3} e^{+2 r} \Big[(x_1-x_2)^2 + (x_2-x_3)^2 + (x_1-x_3)^2\Big]\Big\}.
\nonumber
\end{eqnarray}
Note that this encoding procedure does not require any online
implementations of continuous-variable CNOT gates
\cite{LloydSlotine,Braunstein1}; combining the signal mode
with the offline squeezed ancilla modes at a sequence of beam splitters
is sufficient \cite{Braunstein2}.

Now we send the three modes through individual channels where each channel
acts independently upon {\it every} mode as described by Eq.~(\ref{errormodel})
with $W_{\rm error}$ corresponding to an $x$-displacement by $\bar x_2$.
As a result, the three noisy channels will turn the encoded state into
the following three-mode state,
\begin{eqnarray}\label{totalafterchannel}
&&W_{\rm enc}'(\alpha_1,\alpha_2,\alpha_3) \\
&&=(1-\gamma)^3
W_{\rm enc}(\alpha_1,\alpha_2,\alpha_3)
\nonumber\\
&&\quad + \gamma (1-\gamma)^2
W_{\rm enc}(x_1-\bar x_2 + i p_1,\alpha_2,\alpha_3)
\nonumber\\
&&\quad + \gamma (1-\gamma)^2
W_{\rm enc}(\alpha_1,x_2-\bar x_2 + i p_2,\alpha_3)
\nonumber\\
&&\quad + \gamma (1-\gamma)^2
W_{\rm enc}(\alpha_1,\alpha_2,x_3-\bar x_2 + i p_3)
\nonumber\\
&&\quad + \gamma^2 (1-\gamma)
W_{\rm enc}(x_1-\bar x_2 + i p_1,x_2-\bar x_2 + i p_2,\alpha_3)
\nonumber\\
&&\quad + \gamma^2 (1-\gamma)
W_{\rm enc}(x_1-\bar x_2 + i p_1,\alpha_2,x_3-\bar x_2 + i p_3)
\nonumber\\
&&\quad + \gamma^2 (1-\gamma)
W_{\rm enc}(\alpha_1,x_2-\bar x_2 + i p_2,x_3-\bar x_2 + i p_3)
\nonumber\\
&&\quad + \gamma^3
W_{\rm enc}(x_1-\bar x_2 + i p_1,x_2-\bar x_2 + i p_2,x_3-\bar x_2 + i p_3).
\nonumber
\end{eqnarray}
Note that we assumed the same $x$-displacements in every channel.
Let us now consider the decoding procedure
and how to extract the error syndromes by using homodyne detections
on the ancilla modes.

\section{Decoding and syndrome extraction}

The decoding procedure now simply means inverting the tritter, which results in
\begin{eqnarray}\label{totalafterdecoding}
&&W_{\rm dec}(\alpha_1,\alpha_2,\alpha_3) \\
&&=(1-\gamma)^3
W_{\rm in}(x_1,p_1)W_{\rm anc}(x_2,p_2)W_{\rm anc}(x_3,p_3)
\nonumber\\
&&\quad + \gamma (1-\gamma)^2
W_{\rm in}\left(x_1-\frac{1}{\sqrt{3}}\bar x_2,p_1\right)\nonumber\\
&&\quad\quad\times
W_{\rm anc}\left(x_2-\sqrt{\frac{2}{3}}\bar x_2,p_2\right)
W_{\rm anc}(x_3,p_3)
\nonumber\\
&&\quad + \gamma (1-\gamma)^2
W_{\rm in}\left(x_1-\frac{1}{\sqrt{3}}\bar x_2,p_1\right)\nonumber\\
&&\quad\quad\times
W_{\rm anc}\left(x_2+\frac{1}{\sqrt{6}}\bar x_2,p_2\right)
W_{\rm anc}(x_3-\frac{1}{\sqrt{2}}\bar x_2,p_3)
\nonumber\\
&&\quad + \gamma (1-\gamma)^2
W_{\rm in}\left(x_1-\frac{1}{\sqrt{3}}\bar x_2,p_1\right)\nonumber\\
&&\quad\quad\times
W_{\rm anc}\left(x_2+\frac{1}{\sqrt{6}}\bar x_2,p_2\right)
W_{\rm anc}(x_3+\frac{1}{\sqrt{2}}\bar x_2,p_3)
\nonumber
\end{eqnarray}
\begin{eqnarray}
&&\quad + \gamma^2 (1-\gamma)
W_{\rm in}\left(x_1-\frac{2}{\sqrt{3}}\bar x_2,p_1\right)\nonumber\\
&&\quad\quad\times
W_{\rm anc}\left(x_2-\frac{1}{\sqrt{6}}\bar x_2,p_2\right)
W_{\rm anc}(x_3-\frac{1}{\sqrt{2}}\bar x_2,p_3)
\nonumber\\
&&\quad + \gamma^2 (1-\gamma)
W_{\rm in}\left(x_1-\frac{2}{\sqrt{3}}\bar x_2,p_1\right)\nonumber\\
&&\quad\quad\times
W_{\rm anc}\left(x_2-\frac{1}{\sqrt{6}}\bar x_2,p_2\right)
W_{\rm anc}(x_3+\frac{1}{\sqrt{2}}\bar x_2,p_3)
\nonumber\\
&&\quad + \gamma^2 (1-\gamma)
W_{\rm in}\left(x_1-\frac{2}{\sqrt{3}}\bar x_2,p_1\right)\nonumber\\
&&\quad\quad\times
W_{\rm anc}\left(x_2+\sqrt{\frac{2}{3}}\bar x_2,p_2\right)
W_{\rm anc}(x_3,p_3)
\nonumber\\
&&\quad + \gamma^3
W_{\rm in}\left(x_1-\sqrt{3}\bar x_2,p_1\right)\nonumber\\
&&\quad\quad\times
W_{\rm anc}\left(x_2,p_2\right)
W_{\rm anc}(x_3,p_3).
\nonumber
\end{eqnarray}
By looking at this state, we can easily see that
$x$-homodyne detections of the ancilla modes 2 and 3
(the syndrome measurements) will
almost unambiguously identify in which channel
a displacement error occurred and
how many modes were subject to a displacement error.
The only ambiguity comes from the case of an error
occurring in every channel at the same time (with probability $\gamma^3$),
which is indistinguishable from the case where no error at all happens.
In both cases, the two ancilla modes
are transformed via decoding back into the two initial
single-mode squeezed vacuum states. All the other cases, however,
can be identified, provided the initial squeezing $r$ is sufficiently
large such that the displacements $\propto\bar x_2$, originating from
the errors, can be resolved in the ancilla states.
This will always be
possible when $r\to\infty$ and/or $\bar x_2\gg 1$.
Note that even without squeezing, $r=0$, all those cases can be
distinguished, provided that $\bar x_2\gg 1$ holds. However,
in this case, the recovery displacement operations will
result in larger excess noises. Perfect recovery with unit
fidelity requires infinite squeezing. Similarly, small displacements
$\bar x_2$ would require sufficiently large squeezing to be resolved,
$e^{-2r}/4<\bar x_2$.

The recovery operation, i.e., the final phase-space displacement
of mode 1 depends on the syndrome measurement results for modes 2 and 3
which are consistent with either undisplaced squeezed vacuum states
(`$0$') or squeezed vacua displaced in either `$+$' or `$-$' $x$-direction.
The syndrome results for modes 2 and 3 corresponding to the
eight possibilities for the errors occurring in the three channels
are ($0$,$0$) for no error at all, ($+$,$0$) for an error in channel 1,
($-$,$+$) for an error in channel 2, ($-$,$-$) for an error in channel 3,
($+$,$+$) for errors in channels 1 and 2, ($+$,$-$) for errors in channels 1 and 3,
($-$,$0$) for errors in channels 2 and 3, and, again, ($0$,$0$) for errors
occurring in all three channels.

The (unnormalized)
conditional states of mode 1 depending on the syndrome measurement results $x_2$
and $x_3$, including suitable feedforward operations
(i.e., displacements of mode 1 using the measured results),
can be obtained by integrating
Eq.~(\ref{totalafterdecoding}) over $p_2$ and $p_3$,
\begin{eqnarray}\label{totalafterdecodingsyndromeandfeedforward}
&&W_{\rm cond}(\alpha_1|x_2,x_3) \nonumber\\
&&=(1-\gamma)^3
\sqrt{\frac{2}{\pi e^{-2r}}} e^{-2 e^{+2r} x_2^2}
\sqrt{\frac{2}{\pi e^{-2r}}} e^{-2 e^{+2r} x_3^2}\nonumber\\
&&\quad\quad\times
W_{\rm in}(x_1,p_1)
\nonumber\\
&&\quad + \gamma (1-\gamma)^2
\sqrt{\frac{2}{\pi e^{-2r}}} e^{-2 e^{+2r} (x_2-\sqrt{2/3}\bar x_2)^2}\nonumber\\
&&\quad\quad\times
\sqrt{\frac{2}{\pi e^{-2r}}} e^{-2 e^{+2r} x_3^2}\nonumber\\
&&\quad\quad\times
W_{\rm in}\left(x_1-\frac{1}{\sqrt{3}}\bar x_2+\frac{1}{\sqrt{2}}x_2,p_1\right)
\nonumber\\
&&\quad + \gamma (1-\gamma)^2
\sqrt{\frac{2}{\pi e^{-2r}}} e^{-2 e^{+2r} (x_2+\bar x_2/\sqrt{6})^2}\nonumber\\
&&\quad\quad\times
\sqrt{\frac{2}{\pi e^{-2r}}} e^{-2 e^{+2r} (x_3-\bar x_2/\sqrt{2})^2}\nonumber\\
&&\quad\quad\times
W_{\rm in}\left(x_1-\frac{1}{\sqrt{3}}\bar x_2+\sqrt{\frac{2}{3}}x_3,p_1\right)
\nonumber\\
&&\quad + \gamma (1-\gamma)^2
\sqrt{\frac{2}{\pi e^{-2r}}} e^{-2 e^{+2r} (x_2+\bar x_2/\sqrt{6})^2}\nonumber\\
&&\quad\quad\times
\sqrt{\frac{2}{\pi e^{-2r}}} e^{-2 e^{+2r} (x_3+\bar x_2/\sqrt{2})^2}\nonumber\\
&&\quad\quad\times
W_{\rm in}\left(x_1-\frac{1}{\sqrt{3}}\bar x_2-\sqrt{\frac{2}{3}}x_3,p_1\right)
\nonumber\\
&&\quad + \gamma^2 (1-\gamma)
\sqrt{\frac{2}{\pi e^{-2r}}} e^{-2 e^{+2r} (x_2-\bar x_2/\sqrt{6})^2}\nonumber\\
&&\quad\quad\times
\sqrt{\frac{2}{\pi e^{-2r}}} e^{-2 e^{+2r} (x_3-\bar x_2/\sqrt{2})^2}\nonumber\\
&&\quad\quad\times
W_{\rm in}\left(x_1-\frac{2}{\sqrt{3}}\bar x_2+2\sqrt{\frac{2}{3}}x_3,p_1\right)
\nonumber\\
&&\quad + \gamma^2 (1-\gamma)
\sqrt{\frac{2}{\pi e^{-2r}}} e^{-2 e^{+2r} (x_2-\bar x_2/\sqrt{6})^2}\nonumber\\
&&\quad\quad\times
\sqrt{\frac{2}{\pi e^{-2r}}} e^{-2 e^{+2r} (x_3+\bar x_2/\sqrt{2})^2}\nonumber\\
&&\quad\quad\times
W_{\rm in}\left(x_1-\frac{2}{\sqrt{3}}\bar x_2-2\sqrt{\frac{2}{3}}x_3,p_1\right)
\nonumber\\
&&\quad + \gamma^2 (1-\gamma)
\sqrt{\frac{2}{\pi e^{-2r}}} e^{-2 e^{+2r} (x_2+\sqrt{2/3}\bar x_2)^2}\nonumber\\
&&\quad\quad\times
\sqrt{\frac{2}{\pi e^{-2r}}} e^{-2 e^{+2r} x_3^2}\nonumber\\
&&\quad\quad\times
W_{\rm in}\left(x_1-\frac{2}{\sqrt{3}}\bar x_2-\sqrt{2}x_2,p_1\right)
\nonumber\\
&&\quad + \gamma^3
\sqrt{\frac{2}{\pi e^{-2r}}} e^{-2 e^{+2r} x_2^2}\nonumber\\
&&\quad\quad\times
\sqrt{\frac{2}{\pi e^{-2r}}} e^{-2 e^{+2r} x_3^2}\nonumber\\
&&\quad\quad\times
W_{\rm in}\left(x_1-\sqrt{3}\bar x_2,p_1\right)\,.
\end{eqnarray}
We observe that in almost all cases, the feedforward operations
turn mode 1 back into the initial state up to some Gaussian-distributed excess
noise depending on the degree of squeezing used for the encoding.
The only case for which no correction occurs is when errors appear
in every channel at the same time, at a probability of $\gamma^3$.
In this case, the initial state remains uncorrected, with an $x$-displacement
error of $\sqrt{3}\bar x_2$.

In the limit of infinite squeezing, $r\to\infty$, the Gaussian distribution functions
in Eq.~(\ref{totalafterdecodingsyndromeandfeedforward})
become delta functions. As a result, the ensemble output state of mode 1
upon averaging over all syndrome measurement results $x_2$ and $x_3$ (by integrating
over $x_2$ and $x_3$) becomes
\begin{eqnarray}\label{ensembleoutput}
(1-\gamma^3)W_{\rm in}(x_1,p_1) +
\gamma^3 W_{\rm in}\left(x_1-\sqrt{3}\bar x_2,p_1\right)\,.
\end{eqnarray}
We see that a fidelity of at least $1-\gamma^3$ can be achieved
(assuming $\bar x_2\gg 1$; for small $\bar x_2$, the fidelity
would exceed $1-\gamma^3$, but those smaller $\bar x_2$ may be too
hard to detect at the syndrome extraction, depending on the
degree of squeezing, see below).
This result implies that the encoded scheme performs
better than the unencoded scheme (direct transmission with
$F_{\rm direct}=1-\gamma$) for {\it any} $0<\gamma<1$.
In other words, by employing the quantum error correction protocol,
the error probability can be reduced from
$\gamma$ to $\gamma^3$. The continuous-variable scheme,
in this model, is more efficient than the analogous qubit repetition
code, and it does not require error probabilities $\gamma < 1/2$
as for the case of qubit bit-flip errors \cite{NielsenChuang}.

Consider now the regime $e^{-2r}/4 < \bar x_2 < 1/4$, corresponding
to small displacements below the vacuum limit.
The resulting displacements can only be resolved provided
the squeezing is large enough. In the limit of infinite squeezing
$r\to\infty$, arbitrarily small shifts can be detected and
perfectly corrected (with zero excess noise in the output states
corresponding to unit fidelity). In the regime $\bar x_2\gg 1$,
corresponding to large shifts, even zero squeezing in the ancilla modes
(i.e., vacuum ancilla states) is sufficient for error identification.
For $r=0$ and $\bar x_2\gg 1$, the syndrome measurements still provide enough
information on the location of the error and, to some extent, on the size
of the error. However, for this case of 
``classical error correction'',
the recovery displacements lead to finite excess noises 
in the output state after error correction, originating from the vacuum
ancilla states.

\section{Conclusions}

Using the simple example of a three-wavepacket repetition code, we demonstrated
that for certain stochastic error models, the continuous-variable, 
Gaussian protocol (based on offline squeezing, beam splitter
transformations, and homodyne detection) 
leads to a significant improvement of fidelity even when
the errors occur in every channel. In this case, the errors 
correspond to errors in one variable, e.g.
$x$-displacements or any errors decomposable into
$x$-displacements.
The appropriate error model is reminiscent of the most typical qubit
channels such as a bit-flip channel.
In the continuous-variable regime,
these types of stochastic errors map a Gaussian signal state
into a non-Gaussian state represented by a discrete, incoherent mixture of
the input state with a Gaussian (or a non-Gaussian) state;
thus, circumventing the recent nogo result on Gaussian QECC 
\cite{Nisetqph}.

It turns out that the fidelity gain through
encoding compared to direct transmission without encoding
is larger than that for the analogous qubit scheme.
In fact, for the three-qubit bit-flip repetition
code, the eight-dimensional, physical Hilbert space
can be divided into only four orthogonal, logical qubit subspaces, corresponding
to the four cases of no error at all and a bit flip on any
one of the three qubits (corresponding to two classical syndrome bits). 
In the continuous-variable case,
even those events with errors
occurring on two of the three modes simultaneously
can be unambiguously identified and corrected, 
since more error subspaces are available and a correspondingly 
larger amount of syndrome information.
Possible applications and extensions of the scheme considered here are
stochastic error models with quadratic or even cubic or higher-order
$x$-errors, gain optimizations for the recovery
displacements with finite squeezing, higher-level repetition codes
in one variable, and
non-commuting (``truly quantum'') errors in both $x$ and $p$,
requiring more complex codes than just three modes.


\acknowledgments

The author acknowledges the
Emmy Noether programme of the DFG in Germany.
He also thanks Samuel Braunstein and
Akira Furusawa for useful discussions.



\begin{thebibliography}{99}
\bibitem{footnoteGKP} a non-Gaussian protocol is considered in
Ref.~\cite{GKP}, where a discrete-variable state can be encoded
into a continuous-variable oscillator mode.
\end{thebibliography}
\end{document}